\documentclass[10pt,twocolumn]{article}
\usepackage{latexsym,amsfonts}
\def\NP{\rm NP}

\def\OR{\vee}

\def\goesto{\rightarrow}
\def\implies{\Rightarrow}

\def\N{{\bf N}}

\def\PR{{\rm Pr}}

\def\PUR{{\rm PUR}}

\makeatletter
\newcommand{\singlespacing}{\let\CS=
        \@currsize\renewcommand{\baselinestretch}{1}\tiny\CS}
\newcommand{\singlespacingplus}{\let\CS=
        \@currsize\renewcommand{\baselinestretch}{1.15}\tiny\CS}
\newcommand{\doublespacing}{\let\CS=
        \@currsize\renewcommand{\baselinestretch}{1.75}\tiny\CS}
\newcommand{\draftspacing}{\let\CS=
        \@currsize\renewcommand{\baselinestretch}{2.0}\tiny\CS}

\makeatother

\def\desclabel#1{\bf #1\hfil}
\def\desc{\list{}{%
\labelwidth=\leftmargin
\advance \labelwidth by -\labelsep
\let \makelabel=\desclabel}}

\newtheorem{lemma}{Lemma}[section]
\newtheorem{theo}[lemma]{Theorem}
\newtheorem{cor}{Corollary}
\newtheorem{prop}[lemma]{Proposition}

\newtheorem{obs}{Observation}

\newtheorem{defi}{Definition}
\def\qed{\hfill$\Box$\newline\vspace{5mm}}

\newtheorem{conj}{Hypothesis}

\begin{document}
\thispagestyle{empty}
\pagestyle{empty}
\bibliographystyle{unsrt}

\title{Computational Complexity and Phase Transitions \\ (extended 
abstract\footnote{an extended version will be available shortly as 
\cite{istrate:ccc00:full}.} )}
\author{Gabriel Istrate\\ 
        Center for Nonlinear Studies and CIC-3 Division, \\
        Los Alamos National Laboratory, 
        Mail Stop B258, \\ 
        Los Alamos, NM 87545, U.S.A.\\
         e-mail: istrate@lanl.gov}

\date{}

\maketitle 
\thispagestyle{empty}
\begin{abstract}
Phase transitions in combinatorial problems have
recently been shown \cite{2+p:nature} to be useful in locating ``hard'' 
instances of combinatorial 
problems. The connection between computational complexity and the 
existence 
of phase transitions has been addressed in Statistical Mechanics 
\cite{2+p:nature} 
and Artificial Intelligence \cite{cheeseman-kanefsky-taylor}, but not
studied rigorously.   

We take a first step in this direction by investigating the 
existence of sharp thresholds for the class of {\em generalized 
satisfiability problems}, defined by Schaefer \cite{schaefer-dich}.  
In the case when all constraints have a special clausal form we 
completely 
characterize the generalized satisfiability problems that have a sharp 
threshold. While NP-completeness does {\em not} imply the sharpness of 
the 
threshold, our result suggests that the class of counterexamples is 
rather limited, as all such counterexamples can be predicted, with constant success probability  by a {\em single} procedure. 
\end{abstract}

\section{Introduction} 

Which combinatorial problems have ``hard'' instances? Computational 
Complexity is the main theory that attempts to provide answers to this 
question. But it is not the only one. While the concept of \NP-complete 
problem, as a paradigm 
for ``problem with hard instances'', has permeated a wide range of 
fields, 
from Computational Biology to Economics, it is not usually considered 
extremely relevant by practitioners. This happens because 
\NP-completeness is an overly pessimistic, worst-case, concept, and in fact if we're 
not really careful about the 
random model, ``most'' instances of many \NP-complete problems turn out 
to be
 ``easy''. 

Much insight in locating the regions ``where the really hard instances 
are'' 
has come from an analogy with Statistical Mechanics, in the context of 
{\em phase transitions in combinatorial problems}. Recent studies 
\cite{2+p:nature} have 
shown that a certain type of phase transitions (called {\em first-order 
phase
transitions}) is responsible for the exponential slowdown of many 
natural 
algorithms when run on instances at the transition point. 

A natural, and early stated question is whether there exists any 
connection between computational complexity and the existence of a phase 
transition. Obtaining 
an answer to this question is further complicated by the fact that the 
physicists' and computer 
scientists' concepts of phase transitions are different: the former 
pertains 
to combinatorial optimization, and is called {\em order-disorder phase 
transition}, while the latter applies to decision problems and is 
called {\em threshold property}, more specifically a restricted form of 
threshold property 
called {\em sharp threshold}\footnote{see definition~\ref{thr:def}.}.  
It is this type of phase transitions we're primarily interested in this 
paper. 

The above question has been asked for both types of phase transitions: 
Fu \cite{fu:phase} argued that there should be no connection between 
worst-case computational complexity and the existence of an 
order--disorder phase transition, by showing that an \NP-complete problem, number 
partition, has no order-disorder phase transition (however see 
\cite{mertens:partition} that argues that number partition {\em has} an 
order-disorder phase transition under a different random model). The case of 
decision problems is even more spectacular: in a paper that proved very 
influential in the Artificial Intelligence community 
\cite{cheeseman-kanefsky-taylor}, Cheeseman, Kanefsky and Taylor conjectured that roughly 
the difference between tractable and intractable problems, specifically 
between problems in $P$ and 
\NP-complete problems is that: 
\begin{enumerate}
\item \NP-complete problems have a phase transition (sharp threshold) 
with respect to ``some'' order parameter. 
\item in contrast, problems in P lack such a threshold. 
\end{enumerate} 

Their conjecture was at best wishful thinking. First,
they did not make it precise enough, by specifying what an order 
parameter 
is. Second, they had no evidence supporting such a radical statement. 
In fact, examples of problems in P that {\em do} have a sharp threshold 
with respect to a ``reasonable'' order parameter had already long been 
known 
(for instance the probability that a random graph has a connected 
component 
of at least, say, $n^{3/4}$ vertices, by the classical results of 
Erd\H{o}s and R\'{e}nyi \cite{erdos:renyi}). 

A natural question is {\em whether there is any connection at all 
between computational complexity and the existence of a sharp threshold at 
least for problems that possess 
some ``canonical'' order parameter}. One restriction that entails the 
existence of a canonical order parameter is the very one which was used 
in defining 
 threshold properties: {\em monotonicity} \cite{bollob-thomasson}. 
Clearly the above-mentioned 
example shatters the hope of obtaining a version of (2) even for 
monotonic 
problems. A quick argument shows that even (1) should 
fail: in any polynomial degree there exist both monotone problems that 
have 
(or do not have) sharp thresholds. The intuitive reason is that the 
existence of a sharp threshold is a statistical property, that is not 
affected by modifying a given problem on a set of instances that has 
zero measure. On 
the other hand worst-case complexity is sensitive to such changes. The 
result is formally stated as Proposition~\ref{density} in the Appendix. 

Given the above argument it would seem that the question has been 
answered, 
and that no whatsoever connection exists between the two concepts. 
However the examples constructed in Proposition~\ref{density} are 
rather 
artificial, and the overall proof is reminiscent of Ladner's 
\cite{lad:j:np-incomplete} result on the structure of polynomial degrees: we can 
construct 
a set of the desired complexity by starting with a  certain base set 
and ``tuning-up'' its worst-case complexity on a set that is ``small 
enough'' so that this does not affect the other desirable property of 
the base set, having a sharp/coarse threshold.  The question still 
remains 
whether the result remains true if we only consider problems with a 
certain ``natural'' structure. After all, this is true in the case of 
computational complexity: Schaefer \cite{schaefer-dich}, 
showed that, when restricted to the class of {\em generalized 
satisfiability problems}, the rich structure of polynomial m-degrees derived 
from Ladner's results simplifies to only two degrees, P and the degree of 
\NP-complete problems, and obtained a full characterization of such 
problems.  

\begin{defi}\label{dich}
Let $S=\{R_{1},\ldots, R_{p}\}$, $R_{i}\subset \{0,1\}^{r_{i}}$, be a 
finite 
set of relations. An {\em $S$-formula in $n$ variables} is a finite 
conjunction of {\em clauses}, i.e. expressions of the type 
$R_{j}(x_{j,1},\ldots, x_{j,r_{j}})$, with the variables $x_{j}$ chosen from a fixed 
set of $n$ variables $x_{1}, \ldots, x_{n}$. $SAT(S)$ is the problem of 
deciding whether an arbitrary $S$-formula has a satisfying assignment 
$x_{1}\ldots x_{n}$ (one that makes each clause true).  
\end{defi}

A pleasant feature of Schaefer's framework is that every problem 
$SAT(S)$ 
is monotonic. Clearly, an analog of (2) fails in this case as well: the 
density result Proposition~\ref{density} is still true for one of the 
two polynomial degrees, P, as 2-SAT has a sharp threshold 
\cite{mickgetssome}, while e.g. at-most-$2$-HORN-SAT has a coarse 
threshold \cite{istrate:stoc99}. On the other hand there exists some 
evidence that 
some notion of computational intractability implies the existence of 
a sharp threshold: in his celebrated result on sharp thresholds for 
3-SAT
Friedgut gives an example of a NP-complete graph problem having a 
coarse 
threshold: the property of containing either a triangle or a 
``large'' clique. From a probabilistic standpoint the second part is 
``not important''. Moreover, his characterization theorem implies that 
{\em any} graph theoretic property that fails to have a sharp threshold 
can be well ``approximated'' by a tractable property, the property 
of containing a copy of a fixed graph. Finally, there is an altogether 
different reasons for a rigorous study of sharp thresholds in 
satisfiability 
problems: in this case the notion of a first-order phase transition 
(that, 
as mentioned {\em does have} significant algorithmic implications) has 
a 
nice combinatorial interpretation, as a ``sudden jump'' in the relative 
size of a combinatorial parameter called {\em backbone} (see e.g. 
\cite{scaling:window:2sat} 
for definition and discussion). It is easy to show (this is an argument 
implicitly made in \cite{2+p:nature}, that will be presented in the 
full version of the paper) that the discontinuity of the backbone 
implies
the existence of a sharp threshold. Therefore studying problems with 
sharp 
thresholds is a useful first step towards identifying all satisfiability 
problems having a first order phase transition.  

It is, perhaps, tempting to conjecture that, when restricted to 
Schaefer's 
framework an analogue of (1) holds: 

\begin{conj}\label{conj}
Every generalized satisfiability problem $SAT(S)$ that Schaefer's 
dichotomy 
theorem \cite{schaefer-dich} identifies as \NP-complete has a sharp 
threshold.\end{conj} 

We further restrict our framework to the case when all constraints 
in $S$ have a special, clausal form. In this case we obtain
a complete characterization of all sets of constraints $S$ for which 
$SAT(S)$ has a sharp threshold. In a preliminary version of this paper 
we claimed that for clausal constraints  \NP-completeness implies 
the existence of a sharp threshold. Unfortunately this is not true, 
as the revised version of our result shows. On the other hand, as 
displayed by Corollary ~\ref{corolar}, the 
class of counterexamples is rather limited: they are those NP-complete 
problems for which satisfiability of a random instance $\Phi$ 
can be predicted with significant success by a very trivial heuristic: 
if neither 
$0^{n}$ or $1^{n}$ are satisfying assignments then return 
``unsatisfiable''.  
So the lack of a sharp threshold {\bf does} have algorithmic 
implications, 
albeit in a probabilistic sense.

\section{Preliminaries} 

We will work in the context of {\em \NP --decision problems}, a 
standard concept in Complexity Theory. For a precise definition see, e.g., 
\cite{papad:b:complexity}. 

\begin{defi} 
The \NP-decision problem $P$ is {\em monotonically decreasing} if for 
every 
instance $x$ of $P$ and every
  witness $y$ for $x$, $y$ is a witness for every instance $z$
  obtained by turning some bits of $x$ from 1 to 0. Monotonically
  increasing problems are defined similarly.
\end{defi}

The three main random model from random graph theory, the so-called 
{\em constant probability model}, the {\em counting} and {\em multiset} 
model extend directly to \NP-decision problems, and are interchangeable 
under quite liberal conditions. For technical convenience we will use 
the constant probability model when proving sharp thresholds and the 
multiset models when dealing with 
coarse thresholds. The following is a brief review.  
The multiset model, 
denoted $\Omega(n,m)$, and which  has two integer parameters $n,m$. A 
random
sample from $\Omega(n,m)$ is obtained by starting with the string 
$z=0^{n}$, 
choosing (uniformly at random and with repetition) $m$ bits of $z$, and 
flipping these bits to one. When $n$ is known, we use $\mu_{m}(A)$ to 
refer to the measure of a set $A$ under this random model. The constant 
probability model 
denoted $\Omega_{p}(n)$ has two parameters, an integer $n$ and a real 
number 
$p\in [0,1]$.  A random sample from $\Omega_{p}(n)$ is obtained by 
starting with the string $z=0^{n}$ and then flipping the bits of $z$ to one 
independently with probability $p$. 
 
\begin{defi}\label{thr:def}
Let $P$ be any monotonically decreasing 
decision problem under the constant probability model
$\Omega_{p}(n)$. A function $\overline{\theta}$ is a
{\em threshold function for $P$} 
if for every function $m$, defined on the set of admissible instances
and taking real values, we have
\begin{enumerate}
\item
if $p(n)=o(\overline{\theta}(n))$ then
$\lim_{n\goesto \infty}
\PR_{x\in \Omega_{p}(n)}[x\in P]=1$,
and 
\item
if $p(n)=\omega(\overline{\theta}(n))$ then
$\lim_{n\goesto \infty}
\PR_{x\in \Omega_{p}(n)}[x\in P]=0$. 

$P$ has a {\em sharp threshold} if in addition the following
property holds:

\item For every $\epsilon >0$ define the functions $p_{\epsilon}(n),
  p_{1/2}(n),p_{1-\epsilon}(n)$ by 
\[ \PR_{x\in \Omega_{p_{\epsilon}}(n)}[x\in P]=\epsilon,
\]
 
\[
\PR_{x\in \Omega_{p_{1/2}}(n)}[x\in P]= 1/2\},
\]
\[ \PR_{x\in \Omega_{p_{1-\epsilon}}(n)}[x\in P]=1-\epsilon
\]
 
 Then we have  
\[
\\lim_{n\goesto 
\infty}\frac{p_{1-\epsilon}(n)-p_{\epsilon}(n)}{p_{1/2}(n)}=0.
\] 
\end{enumerate}

If, on the other hand, for some $\epsilon >0$ the amount
$\frac{p_{1-\epsilon}(n)-p_{\epsilon}(n)}{p_{1/2}(n)}$ is bounded away 
from 0
as $n\goesto \infty$,  $P$ has a {\em coarse threshold}. These
two cases are not exhaustive as the above quantity could in principle
oscillate with $n$. Nevertheless they are so for most ``natural''
problems.  
\end{defi}

Let $f:{\bf N}\goesto {\bf R}$. Define $QEMPTY(f)$ to be the 
probability that the following queuing chain: 
\[
\left \{\begin{array}{l}
Q_{0}=1,\\
Q_{i+1}=Q_{i}-1+\Xi_{i+1}.
\end{array}
\right.
\]

(where the $\Xi_t$'s are independent Poisson variables with parameter 
$f(t)$)
ever remains without customers. 

\begin{defi}
Let $(a,b)\in \N \times \N \setminus (0,0)$. Define 
$C_{a,b}=\overline{x}_{1}\OR \ldots \OR \overline{x}_{a}\OR x_{a+1}\OR \ldots \OR 
x_{a+b}$. Such a relation is called {\em clausal constraint}.
\end{defi}

For a set $S$ as in definition~\ref{dich} let $k$ be the maximum arity 
of a 
relation in $S$. To avoid trivial cases, we assume that $k\geq 2$. For 
$i=\overline{1,k}$ 
let $p_{i}$ be 1 if clause $\overline{x}_{1}\OR \ldots \OR 
\overline{x}_{i-1}\OR x_{i}\in S$ and 0 otherwise, and let $n_{i}$ be 1 if clause 
$\overline{x}_{1}\OR \ldots \OR \overline{x}_{i}\in S$ and 0 otherwise. 
Define polynomials $P_{i}(c)=\sum_{j\geq i}{{c}\choose {j-i}}\cdot 
p_{j}$ and $Q_{i}(c)=\sum_{j\geq i}{{c}\choose {j-i}}\cdot n_{j}$. Let 
$\delta_{k}=kp_{k}+n_{k}$, $N_{S}={{n}\choose {k}}\cdot \delta_{k}$, and 
$\alpha =m/N_{S}$.  
Finally, let 
\[
a_{0}=\max \{0\}\cup \{a:C_{a,0}\in S\},
\]
\[
a_{\geq 1}=\max \{0\}\cup \{a:C_{a,b}\in S,b\geq 1\}. 
\]

$b_{0}$ and $b_{\geq 1}$ are defined similarly with respect to the 
second component. 

\section{Main result} 

Recall that a relation is called {\em 0-valid} (1-valid) if it is 
satisfied by the assignment ``all zeros'' (``all ones'') and {\em Horn} 
(negated Horn) if it is equivalent to a Horn (negated Horn) 
CNF-formula. When $S$ is Horn the number of clauses in $S$ over $n$ variables is 
$N_{S}(1+o(1))$.
For a property $T$ we will use ``$S$ is $T$'' as a substitute for 
``every relation in $S$ is $T$''. 

 Our main result is
\begin{theo}\label{main}
Let $S$ be a finite set of {\em clausal constraints}. 
\begin{description}
\item{a. } If $S$ is 0-valid or $S$ is 1-valid then the decision 
problem $SAT(S)$ is trivial. 
\item{b. } If $S$ is (Horn $\cup$ 0-valid) or $S$ is (negated Horn 
$\cup$ 1-valid) then 
$SAT(S)$ has a coarse threshold. 
\item{c. } Suppose  cases $a$. and $b$. do not apply. If 
\[(a_{\geq 1}< a_{0}\leq b_{0}) \OR (b_{\geq 1}< b_{0}\leq 
a_{0})\OR \]
\[ (a_{0}=b_{0}=\min\{a_{\geq 1},b_{\geq 1}\})
\]
then $SAT(S)$ has a sharp threshold, otherwise $SAT(S)$ has a coarse 
threshold.

\end{description}
\end{theo}

For reasons of space we can do little but present a rather sketchy 
outline of the proof of 
Theorem~\ref{main}. A full version will be given in 
\cite{istrate:ccc00:full}.
The following corollary (of the preceding result and its proof) 
summarizes the intuition that all \NP-complete problems with coarse 
thresholds are ``rather trivial''. 
  
\begin{cor}\label{corolar}
Suppose $S$ is a finite set of clausal constraints. Then $SAT(S)$ has a 
coarse threshold {\em exactly} when at least one of the following 
(non-exclusive) conditions applies. 
\begin{enumerate}
\item $S$ is Horn. 
\item $S$ is negated Horn. 
\item $SAT(S)$ is \NP-complete and has the same threshold function as 
the 
property ``$0^{n}$ satisfies $\Phi$''. 
\item $SAT(S)$ is \NP-complete and has the same threshold function as 
the 
property ``$1^{n}$ satisfies $\Phi$''.  
\end{enumerate}
\end{cor}

Indeed, in the cases 3 and 4 there exists {\em a single trivial algorithm}, 
that declares the formula unsatisfiable if it is not satisfied by any of
the two 
assignments $0^{n}$ and $1^{n}$, and which  is correct with a constant 
probability $\epsilon$ over the {\em whole} range of the parameter 
$p$ (in the constant probability model). 

\begin{obs} 
In the general case there are other (non-clausal) examples of 
satisfiability 
problems with a coarse 
threshold. Let $R(x,y)$ be the relation $''x\neq y''$. 
Then $SAT(\{R\})$ is essentially the  2-coloring problem, which has a 
coarse threshold. 
\end{obs}

\section{Proof sketch}
\begin{center}
\begin{figure}
{\tt
\begin{tabbing}

Pr\=ogram PUR($\Phi$): \\
if \= $\Phi$ (contains no positive unit clause)\\
\>\= return TRUE \\
else \\
\>choose such a positive unit clause $x$ \\
\>if \= ($\Phi$ contains $\overline{x}$ as a clause)\\
\> \>return FALSE \\
\> else \\
\> \>let $\Phi^{\prime}$ be the formula \\
\> \>obtained by setting
   $x$ to 1 \\
\> \>return \PUR($\Phi^{'}$) \\
\end{tabbing}
}
\caption{Algorithm PUR}
\end{figure}
\end{center}

\begin{description}
\item{b. } 
This part of the proof is constructive. When $\Phi$ is Horn we 
explicitly determine the probability that a random formula $\Phi$ is 
satisfiable, and then 
use it to argue that the corresponding (Horn $\cup$ 0-valid) cases also 
have a coarse threshold. The analysis of the Horn cases is similar to 
the one 
when $S$ consists of all Horn clauses of length at most $k$, that was 
settled in \cite{istrate:stoc99}, and is accomplished by analyzing PUR, 
a natural implementation of positive unit resolution, which is complete 
for 
Horn satisfiability. 

We regard \PUR\ as working in stages, indexed by the
number of variables still left unassigned; thus, the stage number
decreases as \PUR\ moves on. We say that {\em formula $\Phi$ survives
Stage $t$} if \PUR\ on input $\Phi$ does not halt at Stage $t$ or
earlier. Let $\Phi_i$ be the formula at the
beginning of stage $i$, and let $N_{i}$ denote the number of its
clauses. We will also denote by $P_{i,t} (N_{i,t})$, the number of 
clauses of
$\Phi_{t}$ of size $i$ and containing one (no) positive
literal. Define $\Phi_{i,t}^{P}$ ($\Phi_{i,t}^{N}$) to be the
subformula of $\Phi_{t}$ containing the clauses counted by $P_{i,t} 
(N_{i,t})$.
The analysis proceeds by showing that we can characterize 
the evolution of PUR on a random formula by a Markov chain, and is 
based on the
following ``Uniformity Lemma'' from \cite{istrate:stoc99}, valid in our 
context as well: 
\begin{lemma}\label{k:3etc:recurrence}
Suppose that $\Phi$ survives up to stage $t$. Then, conditional on the
values 
$(P_{1,t}, N_{1,t},\ldots, P_{k,t}, N_{k,t})$, the clauses in  
$\Phi_{1,t}^{P},
\Phi_{1,t}^{N},\ldots,  \Phi_{k,t}^{P}, \Phi_{k,t}^{N}$ are chosen 
uniformly
at random and are independent. Also, conditional on the
fact that $\Phi$ survives stage $t$ as well, the following recurrences
hold:
\[
\left \{\begin{array}{l}
         P_{1,t-1}=P_{1,t}-1-\Delta_{01,t}^{P}+\Delta_{12,t}^{P}, \\        
         N_{1,t-1}=N_{1,t}+\Delta_{12,t}^{N},                    \\
         \end{array}
\right.
\]
and, for $i=\overline{2,k}$,
\[
\left \{\begin{array}{l}
         
P_{i,t-1}=P_{i,t}-\Delta_{0i,t}^{P}-\Delta_{(i-1)i,t}^{P}+\Delta_{i(i+1),t}^{P}, \\        
         N_{i,t-1}=N_{i,t}-\Delta_{(i-1)i,t}^{N}+\Delta_{i(i+1),t}^{N},                    
\\
         \end{array}
\right.
\]
where 
\[
\left \{\begin{array}{l}
\Delta_{01,t}^{P} =B(P_{1,t}-1,1/t),\\
\Delta_{(i-1)i,t}^{P}=B(P_{i,t},(i-1)/t),\\
\Delta_{0i,t}^{P}=B(P_{i,t}-\Delta_{(i-1)i,t}^{P},1/t),\\
\Delta_{(i-1)i,t}^{N}=B(N_{i,t},i/t),\\
\Delta_{k(k+1),t}^{P}=\Delta_{k(k+1),t}^{N}=0.\\
\end{array}
\right.
\]
\end{lemma}

The main intuition for the proof is that with high probability the 
binomial expressions in the previous formulas are close to their expected 
values. The proof of this very intuitive statement is conceptually 
simple, but technically 
somewhat involved, and mirrors the proof in 
\cite{istrate:stoc99}. So all it remains is to characterize the mean 
values
of $P_{i,t}$, $N_{i,t}$. We only outline the main steps of this 
computation in 
the sequel, assuming that the above mentioned concentration results 
hold. 
Define $x_{i,t},y_{i,t}$ by 
\[
\left \{\begin{array}{l}
E[P_{i,t}]=i\cdot {{t}\choose {i}}\cdot x_{i,t},\\
E[N_{i,t}]={{t}\choose {i}}\cdot y_{i,t}.
\end{array}
\right.
\] 
Then it is easy to see that sequences $x_{i,t},y_{i,t}$, $i\geq 2$ 
verify the recurrences: 
\[
\left \{\begin{array}{l}
x_{i,t-1}=x_{i,t}+x_{i+1,t},\\
y_{i,t-1}=y_{i,t}+y_{i+1,t}.
\end{array}
\right.
\] 

Define the vector sequence $(Z_{t})_{t\geq 0}\in {\bf R}^{k-1}$ by 
$Z_{t+1}=A \cdot Z_{t}$, with $A=(a_{i,j})$, 
\[a_{i,j}=\left \{\begin{array}{ll}
                         1, & \mbox{ if }j=i+1,\\
                         0, & \mbox{ otherwise.}
\end{array}
\right.
\]

It is easy to see that both sequences $(x_{i,t})_{t}$ and 
$(y_{i,t})_{t}$ 
satisfy the same recurrence as $Z_{t}$. A simple computation shows that 
$A^{k}_{i,j}= {{k}\choose {j-i}}$ (where, for $t<0$, ${{k}\choose 
{t}}=0$). 
Therefore $Z_{i,t}=\sum_{j\geq i}{{t}\choose {j-i}}Z_{i,0}$. Since 
$x_{i,n}=\alpha\cdot p_{i}\cdot (1+o(1))$, we have that for every constant 
$c>0$, $x_{i,n-c}=\alpha \cdot P_{i}(c)\cdot (1+o(1))$ for every $i\geq 
2$. In the same way $y_{i,n-c}=\alpha \cdot Q_{i}(c)\cdot (1+o(1))$. 

Computing $x_{1,t},y_{1,t}$ (or equivalently $P_{1,t},N_{1,t}$) needs 
some care, and this is where several forms of the threshold result are 
obtained. 
\begin{description}
\item{{\bf Case 1: $\exists j_{1},j_{2}\geq 2$, 
$p_{j_{1}}=n_{j_{2}}=1$. }}
The following is the result in this case: 
\begin{theo}
Let $c>0$, and let $m=c\cdot n^{k-1}$. Then the probability that \PUR\ 
accepts 
$\Phi$ is equal to $QEMPTY(c\cdot \frac{k!}{\delta_{k}}\cdot 
P_{2}(j))$. 
\end{theo}

The proof of the theorem goes along the following lines: 
\begin{enumerate}
\item as long as $P_{1,t}$ is ``small'' (sublinear) $P_{1,t-1}\sim 
P_{1,t}-1 + 
Po(t\cdot x_{2,t})$. This is particularly true in the first $\theta(1)$ 
stages, when $P_{1,t}$ can be approximated by a queue with arrival 
distribution $Po(c\cdot \frac{k!}{\delta_{k}}\cdot P_{2}(n-t))$. This 
explains the form of the 
limit probability. 
\item Also, in the first $\theta(1)$ stages $P_{1,t},N_{1,t}$ 
are ``small'' (approximately constant), so that w.h.p. \PUR\ does not 
reject. 
\item The probability that \PUR\ accepts after the first $\theta(1)$ 
stages is small, since, after these stages $P_{1,t}$ will be large enough 
to make a decrement to $0$ unlikely.
\item At the stages $c=n-\theta(\sqrt{n})$,  $P_{1,t},N_{1,t}$ are 
large enough to guarantee the existence, with nonnegligible probability of 
a variable that 
appears both as a positive and a negative unit clause. 
\end{enumerate}

Let $S$ be now (Horn $\cup$ 0-valid), $S_{H}=S\cap HORN$, let $\Phi$ be 
a random formula and 
$\Phi_{H}$ be its ``Horn part''. That $SAT(S)$ has the same (coarse) 
threshold as $SAT(S_{H})$ follows easily from the following set of 
inequalities: 
\[
 \Pr[\Phi \mbox{ has no positive unit clauses }]\leq
\]
\[
\Pr[\Phi \in SAT]\leq
\Pr[\Phi_{H}\in SAT].
\]

\item{{\bf Case 2: $\exists j_{1}\geq 2$, $p_{j_{1}}=1$ but $\forall 
j\geq 2: n_{j}=0$. }}Then the following holds: 
\begin{theo}
Let $c>0$, and let $m=c\cdot n^{k-1}$. Then the probability that \PUR\ 
accepts 
$\Phi$ is equal to \[e^{-c\cdot \frac{k!}{\delta_{k}}}+(1-e^{-c\cdot 
\frac{k!}{\delta_{k}}})\cdot QEMPTY(c\cdot \frac{k!}{\delta_{k}}\cdot 
P_{2}(j)).\]
\end{theo}

The outline is quite similar to the one of the previous case, with a 
couple of differences. 
\begin{enumerate}
\item Now $N_{1,t}$ no longer grows, but remains equal to $N_{1,n}$ for 
as long as the algorithm does not halt. There exist a nonnegligible 
(and asymptotically equal to $e^{-c\cdot \frac{k!}{\delta_{k}}}$) 
probability that $N_{1,n}=0$. In this case 
$11\ldots 11$ is a satisfying assignment. 
\item In the opposite case the structure of the proof (and conclusion) 
is similar to the one from the Case 1, except that, since $N_{1,t}$ no 
longer grows, 
we have to look up to $\theta(n)$ stages to be sure that the algorithm 
has a nonnegligible probability to reject.
In this case the term $\Delta_{01,t}^{P}$ can no longer be taken to be 
approximately zero. One can, however, get by, by noticing that, at 
those stages where $P_{1,t}$ is $\theta(n)$, the probability that there 
exists a positive unit clause opposite to the negative unit clause 
guaranteed by the condition $N_{1,n}>0$
is approximately constant. Iterating this over a small but unbounded 
number of 
steps allows us to conclude that for every $\epsilon >0$ with 
probability $1-o(1)$ the formula becomes unsatisfiable in one of the first 
$\epsilon\cdot n$ 
stages. Taking $\epsilon$ small enough so that $P_{1,t}$ is still 
nonzero 
after $\epsilon\cdot n$ stages (if \PUR\ hasn't already stopped by this 
time) allows us to derive the same form of the limit probability as in 
case 1.    
\end{enumerate}

The analysis of the (Horn $\cup$ 0-valid) case is similar to the 
previous one. 

\item{{\bf Case 3: $\exists j_{2}\geq 2$ $n_{j_{2}}=1$ but $\forall 
j\geq 2: p_{j}=0$.  }}

In this case the threshold result is 
 \begin{theo}
Let $c>0$, and let $m=c\cdot n^{k-1+\frac{1}{k+1}}$. Then the 
probability that \PUR\ accepts 
$\Phi$ is equal to \[e^{-c^{k+1}\cdot (k!)^{k}}+o(1).\] 
\end{theo}

The main steps of the 
analysis are: 
\begin{enumerate}
\item In this case $P_{1,t}$ is decreasing, but the special form of the 
threshold makes sure that $\Delta_{01,t}^{P}$ can be neglected, so 
$P_{1,t-1}\sim P_{1,t}-1$, and $P_{1,t}\sim P_{1,n}-(n-t)$. 
\item On the other hand $N_{1,t}$ increases and approximately satisfies 
the following recurrence $N_{1,t-1}\sim N_{1,t}+(t-1)\cdot y_{2,t}$, 
where $y_{2,t}$ can be computed as outlined before. 
\item The probability that the positive literal chosen at stage $t$ 
occurs both in positive and negative unit form is approximately 
$1-e^{-\frac{N_{1,t}}{t}}$.
\item The threshold interval is obtained when the probability that the 
algorithm rejects in the last $\theta(1)$ stages becomes roughly 
constant 
(so that the events ``PUR accepts'' and ``PUR rejects'' compete). 
\item A recursive computation yields the final form of the limit 
probability.  
\end{enumerate}
\end{description}

An interesting thing happens when considering the corresponding (Horn 
$\cup$ 
0-valid) case: the threshold interval is no longer the one from the 
corresponding Horn case, but rather mirrors the one in Cases 1 and 2. The 
underlying 
reason is simple: the lower bound is the same as in Cases 1 and 2, the 
probability that $\Phi$ contains no positive unit clause. To show an 
upper bound less than one, consider applying $\PUR$ (which is no longer 
complete) to our 
formula. With some positive probability $\PUR$ will exhaust all the 
positive unit literals (including those created on the way) before 
accepting. 
Since $S$ is not Horn, it contains a clause template with $b\geq 2$ positive
literals. 

Such clauses will result, when the positive unit clauses are exhausted, 
into 
an at least linear number of clauses of the type $C_{0,b}$. Together 
with 
the ``all negative'' clauses these will ensure that w.h.p. (at least 
for 
a big enough constant $c$) the remaining formula is unsatisfiable. 
Thus the probability that $\Phi$ is satisfiable is less than 
$1-\Pr[\PUR\mbox{ exhausts all its positive unit clauses}]-o(1)$. 
The only case left uncovered by this argument is when the only type of 
``all negative'' clauses are the unit clauses, but in this case one 
can apply a similar reasoning by setting the variables appearing in negative 
unit clauses too. 

\item{c. } The argument is based on Friedgut's proof 
\cite{friedgut:k:sat} of the fact that 3-SAT has a sharp threshold, and we assume 
familiarity with the concepts and the methods in this paper. He first shows a 
general result that roughly states that graph (and hypergraph) problems 
that have coarse thresholds have a simple approximation at the 
threshold point. Here is a general and 
cleaner version of this result from J. Bourgain's appendix: 
\begin{prop}\label{bourgain}
Let $A\subset \{0,1\}^{n}$ be a monotone property, and assume say
\[
\epsilon\leq \mu_{p}(A)\leq 1-\epsilon
\]
\[
p\frac{d\mu_{p}(A)}{dp}<C
\]
for some $p=o(1)$ and $C>0$\footnote{such $p$ and $C$ exist, assuming 
that the sharp threshold condition for $A$ fails with respect to 
$\epsilon >0$.} Then there is $\delta=\delta(C)$ such that either 
\begin{equation}\label{first-cond}
\mu_{p}(\{x\in \{0,1\}^{n}| x\supset x^{\prime}\in A,
|x^{\prime}|\leq 10C)>\delta
\end{equation}
or there exists $x^{\prime}\not \in A$ of size $|x^{\prime}|\leq 10C$ 
such that the conditional probability 
\begin{equation}\label{second-cond}
\mu_{p}(x\in A| x\supset x^{\prime})>\frac{1}{2}+\delta.
\end{equation}
\end{prop}

As a sanity check, let us see how this theorem applies to the three 
cases of 
HORN-SAT we have just analyzed. The set $A$  is taken to be 
$\overline{SAT(S)}$.
\begin{itemize}
\item In the first two cases condition~\ref{second-cond} applies, and 
the ``magical'' formula $x^{\prime}$ is simply a fixed unit clause.  
\item In the last case condition~\ref{second-cond} applies. The 
``forbidden 
formula'' $x^{\prime}$ consists of $k$ different unit clauses 
$x_{1},\ldots, x_{k}$, together with the clause $\overline{x}_{1}\OR \ldots \OR 
\overline{x}_{k}$. An unexpected outcome of the analysis is that the 
satisfiability probability of a random formula $\Phi$ {\em coincides 
within $o(1)$} with the probability that $\Phi$ contains no isomorphic copy 
of $x^{\prime}$. 
\end{itemize}

Suppose $S$ is neither (Horn $\cup$ 0-valid) nor (negated Horn $\cup$ 
1-valid)

Then $S$ contains the clauses $C_{a_{0},0}$ 
and $C_{0,b_{0}}$ and $a_{0},b_{0}\geq 2$. Assume w.l.o.g. that $b_{0}\leq a_{0}$. 
According to another theorem of Friedgut (that is rederived by Bourgain 
as Corollary 3), there exists $\gamma \in {\bf Q}$ such that the value $p$ 
from Proposition~\ref{bourgain}  is $\theta(n^{\gamma})$. Therefore the expected 
number of 
copies of the clause $C_{0,b_{0}}$ in a random SAT(S) formula is 
$\theta(n^{\gamma_{1}})$, for some rational number $\gamma_{1}$. 
It is easy to see that $\gamma_{1}\geq 0$. Indeed, suppose otherwise. Then
 the expected number of 
copies 
of $C_{0,b_{0}}$ in $\Phi$ is $o(1)$, so with probability $1-o(1)$ $\Phi$ 
contains no clauses consisting of positive literals only. 
Therefore with probability $1-o(1)$ the assignment $0^{n}$ satisfies 
$\Phi$, 
which is a contradiction.

{\bf Case 1: Suppose $b_{\geq 1}< b_{0}$.}\hspace{5mm}

In this case we want to show that $SAT(S)$ has a sharp threshold. 
A first observation is that $\gamma_{1} >0$. Indeed, suppose $\gamma_{1}=0$ 
and consider the formula $\Xi$ obtained from $\Phi$ in the following manner: delete from each clause of 
$\Phi$ of length at least $b_{0}$ (with probability $1-o(1)$ {\em all} clauses 
of $\Phi$ are like that) $b_{0}-1$ literals chosen as follows: 
\begin{itemize}
\item If the clause has at most $b_{0}-1$ positive literals delete them all; then delete a number of random negative literals, so that in the end we delete 
$b_{0}-1$ literals. 
\item Otherwise delete all but one of the $b_{0}$ positive literals, chosen 
uniformly at random. 
\end{itemize}

It is easy to see that $\Xi\in SAT\implies \Phi \in SAT$. $\Xi$ is a Horn 
formula, falling in the third category (since, by the assumption $b_{1}<b_{0}$ no positive remaining clause has length greater than 1). The formula is {\em 
not} a uniform one (since clauses of the same length are {\em not} do not 
have the same probability of occurrence). However it can be made so, 
while increasing the satisfaction probability, by keeping only a fraction of
the clauses that occur with probability higher than the minimum one among 
clauses of the same length. From b. Case 3 it follows that with probability 
$1-o(1)$ $\Xi$ (therefore $\Phi$) is satisfiable, contradiction. 

We are now in position to outline how to mimic Friedgut's argument to 
show a sharp threshold in our case. Friedgut deals directly with the 
monotone set $A$ of k-DNF formulas that are tautologies, and first 
shows that, assuming that this set does not have a sharp 
threshold it is the alternative~\ref{second-cond} that holds. This is evident 
for K-SAT, but not in our case. Fortunately, we can use some of his argument: 
assuming that the other alternative holds, the critical value would be 
$p=\theta(n^{-v/c})$, deriving from an unsatisfiable formula $F$ 
with $v$ variables and $c$ clauses. To give this threshold, $F$ is also {\em balanced}, that is, has ratio clauses/variables higher than any of its induced subformulas. Since $F$ is unsatisfiable it immediately follows that $v<c$. 
But this cannot happen, 
since a first moment method easily shows that in our case $p=o(1/n)$.

He then proceeds to show that for k-SAT 
there cannot exist a ``magical'' formula 
$x^{\prime}$ with the properties guaranteed by Proposition~\ref{bourgain}. 
The proof follows the following outline
(the quotes below refer to statements in \cite{friedgut:k:sat})
\begin{enumerate}
\item the nonexistence of a sharp threshold implies the existence of a 
small ``magical'' formula $F$, which is not itself a tautology, and which 
boosts the probability that a random formula $\Phi$ is a tautology, if 
we condition on $\Phi$ 
containing a {\em fixed} copy of $F$ by a non-negligible ($\Omega(1)$) 
amount. 
\item the existence of such a formula implies that adding a 
constant number of random clauses of size 1 to a random formula also boosts 
the probability of obtaining a tautology by a positive amount. 
\item finally, a contradiction is obtained by showing that were the 
conclusion of the previous step true, then adding instead an arbitrarily  
small (but unbounded) number of clauses of size $k$ would also be 
enough to boost the probability of obtaining a tautology.
But such a statement can be refuted directly (Lemma 5.6). 
\end{enumerate}

The heart of Friedgut's proof is Step 3, 
a geometric argument, Lemma 5.7 in his paper. This is where the 
special syntactical nature of k-SAT (or rather, dually, k-DNF-TAUTOLOGY) 
appears: according to Lemma 5.7, the probability that an arbitrary subset 
of the hypercube $\{0,1\}^{n}$ 
can be covered with a small (but nonconstant) number of hyperplanes of 
codimension $k$ (corresponding to DNF-clauses of length exactly $k$) is 
asymptotically no smaller than the probability that it can be covered 
with a constant number of hyperplanes of codimension 1, whose existence 
is implied by Proposition~\ref{bourgain}
via the process outlined in steps 1,2,3. The clausal structure of 
$k-SAT$ is 
reflected by the correspondence between clauses of size $k$ and 
hyperplanes of codimension $k$, and this correspondence will extend in our 
more general case. The argument in Lemma 5.7 is not specific to 
$k-SAT$, but works in some other cases, if we replace, of course, 
hyperplanes of codimension $k$ by the corresponding type of hyperplanes and 
make sure that the geometric argument still works.
For instance one can mimic the proof to show that $SAT(S_{0})$, where 
$S_{0}=\{C_{a_{0},0},C_{0,b_{0}}\}$ has a sharp threshold.  A minor 
technical nuisance is that now we need to consider two types of 
hyperplanes of codimension larger than one, corresponding 
to both types of clauses, but this does not influence the overall 
reasoning.  

The idea of our argument is now rather transparent: the rest of the 
steps in Friedgut's argument extend more or less in a straightforward fashion, 
and it is only the analog of Lemma 5.7 where we need to see how the 
proof extends. In our case we have a ``large'' (non-constant) number of 
copies of $C_{a_{0},0},C_{0,b_{0}}$ in a random $SAT(S)$ formula 
at the critical value of $p$. They are used to ``cover a finite number of 
unit clauses''. But this property {\em does not depend on the other types of 
clauses in $S$}, as long as we can make sure that we have a non-constant 
number of copies of $C_{a_{0},0},C_{0,b_{0}}$ (this is where $\gamma_{1}>0$ 
comes into play).  

These two types of clauses act as a ``$SAT(S_{0})$'' core of the 
formula $\Phi$, that is enough to ensure that a the geometric argument 
used to prove that $SAT(S_{0})$ has a 
sharp threshold holds for $SAT(S)$ as well. The structure of the 
proof in this case
is similar, at a very high level, with the one of Schaefer's dichotomy 
theorem: in this latter case the canonical problem is 3-SAT 
and \NP-completeness follows from the ability to ``simulate'' all 
clauses of length 3. For sharp/coarse thresholds, the canonical problem is 
$SAT(S_{0})$, and the existence of a sharp threshold follows from the 
ability to ``simulate'' both clauses in $S_{0}$. 
{\bf Case 2: Suppose $a_{0}=b_{0}=b_{\geq 1}\leq a_{\geq 1}$.}\hspace{5mm}

The ideea is similar to the one in Case 1: we show first that the 
expected number of copies of $C_{a_{0},0}$ and $C_{0,b_{0}}$ is not constant 
in the critical region, and use Friedgut's argument for $S_{0}$. The deletion 
process is almost identical to the one of the previous section, except that, 
in order to avoid creating ``all negative'' clauses of length greater than 1, 
we do {\em not} delete the last positive literal, in a clause with less than 
$b_{0}$ positive literals, but a random negative literal.

{\bf Case 3.}\hspace{5mm}

Assume that we are not into either Case 1 or Case 2 
because of the similar inequality 
for $a_{0}$. In this case we want to show that $SAT(S)$ has a coarse 
threshold, occurring for $p$ such that the expected number of copies of 
$C_{0,b_{0}}$ is a constant $c$. We have already seen that the probability 
that a random formula $\Phi$ is satisfiable is lower bounded by the probability that it contains no copies of $C_{0,b_{0}}$. So we only need to argue that the 
satisfaction probability is strictly less than 1, for some high enough value 
of the constant in the definition of $p$. 

The main ingredient of this proof, presented in full in the final version of
the paper, is the claim that resolution will create the empty clause (thus 
certifying that the formula is unsatisfiable) with probability bounded away 
from 0. 
This is easy to see if $a_{0}>b_{0}$ and $b_{\geq 1}> b_{0}$: consider first the set of 
all variables that appear in a copy of $C_{0,b_{0}}$ in $\Phi$ (the number of such clauses has a Poisson 
distribution). The variables in these clauses are different with probability $1-o(1)$. \
A satisfying assignment (if it exists) must satisfy at least one such variable from 
each clause. Choose one variable from each such clause (there are, on the average, a 
constant number of 
ways to do this) and replace each clause by the positive unit clause consisting of 
the chosen variable. If the original formula was satisfiable then the new one is too, 
for at least one choice, corresponding to a satisfying assignment.  

Let us consider the clauses of type $C_{a,b_{1}}$ (with $a\geq 1$ minimal) 
whose negative literals involve chosen variables only, and whose positive literals do {\em not} appear in the copies 
of $C_{0,b_{0}}$. When the number of copies of $C_{0,b_{0}}$ is at least $a$ (which happens 
with probability bounded away from zero) resolution, applied to the new formula, will create a 
number of copies of $C_{0,b_{1}}$ with average $\Omega(n)$ (since $b_{\geq 1}>b_{0}$).  W.h.p. the 
number of such clauses is close to its expected value. Consider now the new clauses of type $C_{0,b_{\geq 1}}$ together with the initial clauses of 
type $C_{a_{0},0}$. With probability $1-o(1)$ (if the constant in the $\theta(1)$ factor 
in $p$ is large enough) this formula is unsatisfiable. Thus resolution will succeed 
with probability bounded away from zero. 
A similar argument (but working with both 
positive and negative variables) works for the case $a_{0}=b_{0}<\min \{a_{\geq 1},b_{\geq 1}\}$. 

The only other remaining case is $b_{\geq 1}=b_{0}<a_{0}$.
Its analysis is slightly more involved, but relies on the same idea: we create 
a linear number of copies of $C_{0,b_{0}}$ by resolving all negative literals 
from copies of $C_{a,b_{\geq 1}}$. The number of copies of $C_{0,b_{0}}$ at each phase is stochastically 
larger than the number of customers in a queuing chain with more clients arriving 
at each stage than those that are served, hence with constant probability it becomes linear. 
Moreover, since only $a$ of the chosen literals can appear negatively, the growth is 
substantially faster than the one of the corresponding queuing chain, in particular the 
number of copies of $C_{0,b_{0}}$ becomes liniar after at most $n^{o(1)}$ iterations of 
the process. In this case the resulting formula is also unsatisfiable with probability $1-o(1)$. 
So the conclusion is the same, that 
the satisfaction probability of a random formula is (for large enough $c$) strictly less than one.  

\end{description}
\qed
\section{Conclusions}

We have investigated the connection between worst-case complexity and 
the existence of phase transitions. Our result shows that some 

connection between the two concepts exists after all: while it is not as clean 
as the one hoped for in \cite{cheeseman-kanefsky-taylor}, the lack  of a 
phase transition has significant computational implications: such problems are either computationally tractable, or well-predicted by 
a single, trivial algorithm. 

Several open problems remain: a first one is to extend our result 
to the whole class of generalized satisfiability problems. 
We believe that obtaining such a characterization is interesting even though 
the motivating conjecture isn't true. Another question is 
whether we can extend apply our techniques to constraint programming problems
(i.e satisfiability over non-binary domains). Obtaining a complete version of 
Schaefer's dichotomy theorem in this case is still open; however 
we believe that some of our results should carry over. 

A third, perhaps the most interesting, open question is to elucidate
 the connection between computational complexity and the ``physical'' 
concept of {\em first-order phase transition}. As we have mentioned, the 
class of problems with such phase transitions is a subset of the class of 
problems with sharp thresholds. For clausal generalized satisfiability 
problems the inclusion is strict: Bollob\'{a}s et 
al. \cite{scaling:window:2sat} have shown that the phase transition in 2-SAT
is of second-order. The proof can perhaps be adapted for any (nontrivial) 
clausal version of 2-SAT. It is tempting to conjecture that at least in the 
clausal case these are all such examples. The non-clausal case is bound to be 
substantially more complex: work in progress \cite{1-in-k:threshold} suggests that 
there exists a (non-clausal) NP-complete generalized satisfiability problem 
with {\em the same width of the scaling window } (and order of the phase transition) as 2-SAT. 
Obtaining any further results is an 
interesting challenge.

\bibliography{bibtheory}
\section*{Appendix}
\begin{prop}\label{density} For every polynomial time degree ${\cal D}$ 
there exist monotone 
\NP-decision problems $A,B\in {\cal D}$ such that 
\begin{itemize}
\item $A$ has a coarse threshold. 
\item $B$ has a sharp threshold. 
\end{itemize}
\end{prop}

{\bf Proof sketch:} 
Start with two problems $C,D\in P$ that have a coarse (sharp) 
threshold, for concreteness the property that a graph contains a triangle and 
2-UNSAT, respectively). Let $E\in {\cal D}$. Encode $E$ into a 
monotonically increasing set $F$ such that $E\equiv_{m}^{P} F$ and 
$\mu_{p}(F)\goesto 1$ as $n\goesto \infty$ for every $p$ in the ``critical region'' 
of $C$.  Define the set $A$ to be the 
set $C\circ F=\{xy| x\in C,y\in F, |x|=|y|\}$. It is easy too see that 
$\mu_{p}(A)=\mu_{p}(C)(1+o(1))$, so $A$ has a coarse threshold. 
Moreover $A\in {\cal D}$. Set $B$ is constructed in a similar fashion. 
\qed
\end{document}